\documentclass[article,aps,nofootinbib,showpacs,amsfonts,epsf]{revtex4}

\input{epsf.tex}

\newcommand{\E}{{\cal{E}}}

\newcommand{\s}{\sigma}

\renewcommand{\a}{\alpha}

\newcommand{\be}{\begin{equation}}
\newcommand{\ee}{\end{equation}}
\newcommand{\bea}{\begin{eqnarray}}
\newcommand{\eea}{\end{eqnarray}}
\newcommand{\ba}{\begin{array}}
\newcommand{\ea}{\end{array}}
\def\J#1#2#3#4{{#1} {\bf #2}, #3 (#4)}
\def\PRD{Phys. Rev. D}
\def\PR{Phys. Rev.}
\def\PRL{Phys. Rev. Lett.}
\def\PTP{Prog. Theor. Phys.}
\def\APN{Ann. Phys. (NY)}
\def\ANNAS{Ann. New York Acad. Scien.}

\def\AP{Ann. Physik}

\def\JGP{J. Geom. Phys.}

\def\JHEP{J. High Energy Phys.}

\def\CQG{Class. Quantum Grav.}
\def\JPA{J. Phys. A: Math. Gen.}

\def\GRG{Gen. Relativ. Grav.}
\def\PLA{Phys. Lett. A}

\begin{document}
\draft

\title{Metric of two balancing Kerr particles\\ in physical parametrization}

\author{V. S.~Manko$^\dag$ and E.~Ruiz$\,^\ddag$}
\address{$^\dag$Departamento de F\'\i sica, Centro de Investigaci\'on y de
Estudios Avanzados del IPN, A.P. 14-740, 07000 M\'exico D.F.,
Mexico\\ $^\ddag$Instituto Universitario de F\'{i}sica Fundamental
y Matem\'aticas, Universidad de Salamanca, 37008 Salamanca, Spain}

\begin{abstract}
The present paper aims at elaborating a completely physical
representation for the general 4-parameter family of the extended
double-Kerr spacetimes describing two spinning sources in
gravitational equilibrium. This involved problem is solved in a
concise analytical form by using the individual Komar masses and
angular momenta as arbitrary parameters, and the simplest
equatorially symmetric specialization of the general expressions
obtained by us yields the physical representation for the
well-known Dietz-Hoenselaers superextreme case of two balancing
identical Kerr constituents. The existence of the physically
meaningful ``black hole-superextreme object'' equilibrium
configurations permitted by the general solution may be considered
as a clear indication that the spin-spin repulsion force might
actually be by far stronger than expected earlier, when only the
balance between two superextreme Kerr sources was thought
possible. We also present the explicit analytical formulas
relating the equilibrium states in the double-Kerr and
double-Reissner-Nordstr\"om configurations.
\end{abstract}


\maketitle


\section{Introduction}

The well-known double-Kerr solution \cite{KNe} was discovered
three and a half decades ago by Kramer and Neugebauer as a
non-trivial application to Einstein's equations of the modern
solution generating techniques in the form of B\"acklund
transformations \cite{Neu}. It gave the researchers an attractive
possibility to study binary systems of interacting Kerr black
holes \cite{Ker}, and in particular to answer an important
question of whether the gravitational attraction of the rotating
black holes can be counterbalanced by their spin-spin interaction.
The equilibrium conditions were first worked out by Kihara and
Tomimatsu \cite{KTo,TKi}, and later Tomimatsu \cite{Tom} obtained
the expressions for the individual Komar \cite{Kom} masses and
angular momenta of the constituents in a generic double-Kerr
configuration. Restricted to the subextreme case, the algebraic
system of the balance equations was solved analytically by
Hoenselaers \cite{Hoe}, who also conjectured, after analyzing
numerically the formulas of Komar masses, that equilibrium between
two Kerr black holes endowed with positive masses is impossible.
At this point, it should be noted that the parametrization
employed in \cite{KNe} does not describe configurations involving
superextreme Kerr constituents, and that is why, for being able to
consider a system of two identical superextreme Kerr sources,
Dietz and Hoenselaers \cite{DHo} used a special complex trick to
pass from the sub- to the superextreme case. Remarkably, they were
able to demonstrate analytically that such a pair of
super-spinning Kerr constituents with positive Komar masses can be
in stationary equilibrium.

Two decades after the publication of the paper \cite{KNe}, a
unified description of the binary equilibrium configurations
composed of arbitrary combinations of the subextreme and
superextreme Kerr constituents became possible due to the
so-called extended double-Kerr (EDK) solution \cite{MRS}
constructed with the aid of Sibgatullin's integral method
\cite{Sib,MSi}. The set of parameters used in the paper \cite{MRS}
turned out to be very advantageous not only for solving
analytically the equilibrium conditions in the general case, which
led in particular to the discovery of the physically meaningful
`subextreme-superextreme' equilibrium configurations, but also for
giving a rigorous proof \cite{MRu1} to Hoenselaers' conjecture on
the non-existence of balance between two black-hole Kerr
constituents with positive Komar masses. Moreover, in recent years
there has been a renewed interest in the double-Kerr solution,
mostly related to the issues of the black-hole configurations with
struts and the geometrical inequalities for black holes. With
regard to the former issue, the research has been principally
directed to the study of the physical properties of two
interacting Kerr black holes \cite{HRe,HRR,MRRS}, while the latter
issue gave birth to a series of papers by Neugebauer and Hennig
\cite{NHe1,HNe,NHe2} in which the aforementioned non-existence
proof \cite{MRu1} was re-examined on the basis of the
area--angular-momentum inequality \cite{HAC} (the validity of this
inequality in the multiple-black-hole case has been proven by
Chru\'sciel et al. \cite{CEN}). It should be emphasized that the
Neugebauer-Hennig analysis, which employs our solution of the
equilibrium problem \cite{MRS}, is in full agreement with the
earlier non-existence proof \cite{MRu1}: the balance of two Kerr
black holes with positive Komar masses is impossible, while a
subextreme constituent with negative mass is unphysical (it
develops a massless ring singularity outside the symmetry axis).

Curiously, although the general equilibrium problem for the EDK
solution was solved more than a decade ago \cite{MRS,MRu1,MRu2},
the physical parametrization of the 4-parameter family of
equilibrium configurations in terms of the Komar quantities has
not yet been obtained up to date. This can be explained by
numerous technical difficulties that one has to overcome for being
able to express all the ``canonical'' parameters of the EDK
solution and various associated constant quantities in terms of
the physical parameters. Recently, nonetheless, we have succeeded
in finding the desired reparametrization  for a 3-parameter
equilibrium configuration \cite{MRu3} that describes a
Schwarzschild black hole levitating in the field of a superextreme
Kerr source, and have studied physical effects in that binary
system. To reach a more ambitious goal, in the present paper we
are going to reparametrize the entire 4-parameter family of the
EDK equilibrium configurations in terms of the Komar physical
quantities. This will be done with the aid of two sets of the
inversion formulas involving parameters of the solution and
individual physical characteristics of the constituents.

The rest of the paper is organized as follows. In the next section
we briefly review the EDK equilibrium configurations and present
the first set of inversion formulas. The Komar individual
characteristics of the constituents and their relation to the
``canonical'' parameters of the EDK solution are discussed in
section 3. The general 4-parameter family of equilibrium binary
systems determined by the EDK solution is reparametrized in
physical parameters in section~4, and the reparametrized
quantities $\s_u$ and $\s_d$ play a crucial role in this process;
here, in particular, we obtain a physical representation for the
Dietz-Hoenselaers (DH) solution \cite{DHo} describing two
identical corotating superextreme Kerr sources in equilibrium. In
section 5 we give a simple new proof of the absence of balance
between two Kerr black holes, and also derive explicit analytical
formulas relating the equilibrium states in the EDK and
double-Reissner-Nordstr\"om \cite{Man} solutions. Section 6
presents concluding remarks.

\section{Solution of the EDK equilibrium problem
in ``canonical'' parameters and the first set of inversion
formulas}

The main advantage of the EDK solution over the non-extended one
originally obtained by Kramer and Neugebauer for two black-hole
constituents \cite{KNe} consists in a remarkable possibility of
its use for solving in a unified manner the equilibrium problem
for any combination of two Kerr sources -- black holes or
superextreme objects. Such a possibility becomes feasible due to
the presence in the EDK solution of the parameters $\a_i$ which
can assume arbitrary real values or occur in complex conjugate
pairs. A pair of two real $\a_i$ then naturally determines an
underextreme Kerr constituent (a black hole if its mass is
positive), while a complex conjugate pair defines a superextreme
constituent (the four main types of binary configurations are
shown in Fig.~1).

The equilibrium configurations in the EDK solution are defined by
an Ernst complex potential $\E$ \cite{Ern} of the following form
\cite{MRu1}:
\bea && \E=\frac{\Lambda+\Gamma}{\Lambda-\Gamma}, \quad
\Lambda=\sum_{1\leq i<j\leq4}\lambda_{ij}r_ir_j, \quad
\Gamma=\sum_{i=1}^4\gamma_ir_i,\nonumber\\
&& \lambda_{ij}=(-1)^{i+j}(\a_i-\a_j)(\a_{i'}-\a_{j'})X_iX_j, \qquad
(i',j'\neq i,j;\,\, i'<j') \nonumber\\
&& \gamma_{i}=(-1)^{i}(\a_{i'}-\a_{j'})(\a_{i'}-\a_{k'})
(\a_{j'}-\a_{k'})X_i, \qquad (i',j',k'\neq i;\,\, i'<j'<k')
\nonumber\\ && r_i=\sqrt{\rho^2+(z-\a_i)^2}, \label{E_pot} \eea
where the parameters $\a_i$, $i=1,2,3,4$, as was already
mentioned, occur as arbitrary real constants or complex conjugate
pairs, and $X_i$ are given by the formulas
\bea &&
X_1=\varphi\frac{\epsilon_1\omega_1-\varphi}{1-\epsilon_1\omega_1\varphi},
\quad
X_2=\varphi\frac{1-\epsilon_1\omega_1\varphi}{\epsilon_1\omega_1-\varphi},
\quad
X_3=-\varphi\frac{1+i\epsilon_4\omega_4\varphi}{i\epsilon_4\omega_4-\varphi},
\quad
X_4=\varphi\frac{i\epsilon_4\omega_4-\varphi}{1+i\epsilon_4\omega_4\varphi}, \nonumber\\
&& \omega_1=\sqrt{\frac{(\a_1-\a_3)(\a_1-\a_4)}
{(\a_2-\a_3)(\a_2-\a_4)}}, \quad \omega_4=\sqrt{\frac{(\a_1-\a_4)(\a_2-\a_4)}
{(\a_1-\a_3)(\a_2-\a_3)}},
\label{Xi} \eea
the complex constant $\varphi$ being subject to the constraint
$|\varphi|^2\equiv\varphi\bar\varphi=1$ (a bar over a symbol means
complex conjugation), while $\epsilon_1=\pm1$ and
$\epsilon_4=\pm1$.

The potential $\E$ defined by (\ref{E_pot})-(\ref{Xi}) is an exact
solution of the Ernst equation \cite{Ern} obtained via
Sibgatullin's method, and the entire metric associated with this
potential has the form \cite{MRS,MRu4}
\bea && d s^2=f^{-1}[e^{2\gamma}(d\rho^2+d z^2)+\rho^2
d\varphi^2]-f(d t-\omega d\varphi)^2, \nonumber\\ &&
f=\frac{\Lambda\bar\Lambda-\Gamma\bar\Gamma}{(\Lambda-\Gamma)(\bar\Lambda-\bar\Gamma)},
\quad
e^{2\gamma}=\frac{\Lambda\bar\Lambda-\Gamma\bar\Gamma}{\lambda_0\bar\lambda_0
r_1r_2r_3r_4}, \quad \omega=2{\rm Im}(\s_0) -\frac{2{\rm
Im}[(G(\bar\Lambda-\bar\Gamma)]}{\Lambda\bar\Lambda-\Gamma\bar\Gamma},
\nonumber\\ && G=z\Gamma+\sum_{1\leq
i<j\leq4}(\a_i+\a_j)\lambda_{ij}r_ir_j
-\sum_{i=1}^4(\a_{i'}+\a_{j'}+\a_{k'})\gamma_ir_i, \nonumber\\ &&
\lambda_0=\sum_{1\leq i<j\leq4}\lambda_{ij}, \quad
\gamma_0=\sum_{i=1}^4\gamma_i, \quad
\s_0=\frac{1}{\lambda_0}[\gamma_0+\sum_{1\leq
i<j\leq4}(\a_i+\a_j)\lambda_{ij}]. \label{metric1} \eea
Mention that the Weyl-Papapetrou cylindrical coordinates $\rho$
and $z$ enter into the potential $\E$ from (\ref{E_pot}) and into
the metric coefficients $f$, $\gamma$, $\omega$ from
(\ref{metric1}) only through the functions $r_i$.

Formulas (\ref{E_pot})-(\ref{metric1}) represent a ``canonical''
form of the solution describing equilibrium configurations in the
EDK spacetime. In order to rewrite them in physical parameters, we
find it helpful first to express the parameters $\a_i$ in terms of
the quantities $\omega_1$ and $\omega_4$. For this purpose we
introduce two additional constants, $z_0$ and $s$, defined as
\be z_0\equiv\frac{1}{4}(\a_1+\a_2+\a_3+\a_4), \quad
s\equiv\frac{1}{2}(\a_1+\a_2-\a_3-\a_4), \label{z0s} \ee
the constant $z_0$ permitting one to make an appropriate choice of
the origin of coordinates on the symmetry axis, and $s$ being the
relative coordinate distance between the centers of the two
constituents.

The sets of $\a$'s describing each type of the binary system in
Fig.~1 are the following (the notation is obvious):
\bea && A_{BB}=\{\a_1>\a_2>\a_3>\a_4\}, \nonumber\\
&& A_{BS}=\{\a_1>\a_2>{\rm Re}(\a_3)={\rm Re}(\a_4),{\rm Im}(\a_3)<0,\a_4=\bar\a_3\}, \nonumber\\
&& A_{SB}=\{{\rm Re}(\a_1)={\rm Re}(\a_2)>\a_3>\a_4,{\rm Im}(\a_1)<0,\a_2=\bar\a_1\}, \nonumber\\
&& A_{SS}=\{{\rm Re}(\a_1)={\rm Re}(\a_2)>{\rm Re}(\a_3)={\rm Re}(\a_4), \nonumber\\
&& \hspace{4cm} {\rm Im}(\a_1)<0,{\rm Im}(\a_3)<0,\a_2=\bar\a_1,\a_4=\bar\a_3\}. \label{Abs} \eea
The proposed change of parametrization is going to transform the
above sets into the new ones, namely,
\bea &&A_{BB} \longrightarrow \Omega_{(0,0)}, \nonumber\\
&&A_{BS} \longrightarrow \Omega_{(0,-1)}\cup\Omega_{(0,+1)}, \nonumber\\
&&A_{SB} \longrightarrow \Omega_{(-1,0)}\cup\Omega_{(+1,0)}, \nonumber\\
&&A_{SS} \longrightarrow \Omega_{(-1,-1)}\cup\Omega_{(+1,-1)}\cup\Omega_{(-1,+1)}, \label{Tr1} \eea
where
\bea && \Omega_{(0,0)}=\{\omega_1>1,\omega_4>1\}, \nonumber\\
&& \Omega_{(0,-1)}=\{\omega_1>1,\omega_4\bar\omega_4=1,{\rm Im}(\omega_4)<0,{\rm Re}(\omega_4)>0\}, \nonumber\\
&& \Omega_{(0,+1)}=\{\omega_1>1,\omega_4\bar\omega_4=1,{\rm Im}(\omega_4)>0,1/\omega_1>{\rm Re}(\omega_4)\ge0\}, \nonumber\\
&& \Omega_{(-1,0)}=\{\omega_4>1,\omega_1\bar\omega_1=1,{\rm Im}(\omega_1)<0,{\rm Re}(\omega_1)>0\}, \nonumber\\
&& \Omega_{(+1,0)}=\{\omega_4>1,\omega_1\bar\omega_1=1,{\rm Im}(\omega_1)>0,1/\omega_4>{\rm Re}(\omega_1)\ge0\}, \nonumber\\
&& \Omega_{(-1,-1)}=\{\omega_1\bar\omega_1=\omega_4\bar\omega_4=1, {\rm Im}(\omega_1)<0,{\rm Im}(\omega_4)<0,{\rm Re}(\omega_1)>0,{\rm Re}(\omega_4)>0\}, \nonumber\\
&& \Omega_{(+1,-1)}=\{\omega_1\bar\omega_1=\omega_4\bar\omega_4=1, {\rm Im}(\omega_1)>0,{\rm Im}(\omega_4)<0,{\rm Re}(\omega_4)>{\rm Re}(\omega_1)\ge0\}, \nonumber\\
&& \Omega_{(-1,+1)}=\{\omega_1\bar\omega_1=\omega_4\bar\omega_4=1, {\rm Im}(\omega_1)<0,{\rm Im}(\omega_4)>0,{\rm Re}(\omega_1)>{\rm Re}(\omega_4)\ge0\}, \label{Om} \eea
and also
\be -\infty<z_0<+\infty, \quad s>0 \label{range_z0} \ee
for all $\Omega$'s. Note that the subindexes in $\Omega$'s have
been designed in such a way that they provide one with the
information about the presence of a black-hole constituent (0) and
the sign of the imaginary part of $\omega_1$ or $\omega_4$.

The inverse parameter change, i.e. the one that maps $\Omega$'s
into the original $A$'s, can be described by means of the
following bi-valued relations ($\delta=\pm1$):
\bea \a_1=z_0+\frac{s}{2}+s\frac{\delta\omega_4(\omega_1^2-1)}{(\omega_1+\delta\omega_4)(1+\delta\omega_1\omega_4)}, \nonumber\\
\a_2=z_0+\frac{s}{2}-s\frac{\delta\omega_4(\omega_1^2-1)}{(\omega_1+\delta\omega_4)(1+\delta\omega_1\omega_4)}, \nonumber\\
\a_3=z_0-\frac{s}{2}+s\frac{\omega_1(\omega_4^2-1)}{(\omega_1+\delta\omega_4)(1+\delta\omega_1\omega_4)}, \nonumber\\
\a_4=z_0-\frac{s}{2}-s\frac{\omega_1(\omega_4^2-1)}{(\omega_1+\delta\omega_4)(1+\delta\omega_1\omega_4)}. \label{Tr1_inv} \eea
It is of course understood that for each $\Omega$ one has to use
only one of the two branches in the above formulas, and the
criterion of choosing the appropriate branch is very simple: if
one of the two subindexes of an $\Omega$ is equal to $+1$ (the
imaginary part of any of the two $\Omega$'s is positive) then one
has to use the branch $\delta=-1$, if not -- then the branch
$\delta=+1$.

We now turn to the consideration of the physical Komar quantities
associated with the EDK solution.

\section{Komar masses and angular momenta. The second set
of inversion formulas}

Explicit analytical formulas for the physical masses and angular
momenta of the balancing constituents in the EDK solution were
obtained in the paper \cite{MRu1}. The Komar masses $m_u$ and
$m_d$ (the subindexes ``u'' and ``d'' are abbreviations from
``up'' and ``down'', referring to the location of the upper and
lower constituents on the symmetry axis) are given by the formulas
\bea
&&m_u=-s\frac{C(C_1-C)}{CC_1+SC_4-1+\epsilon\delta CS}, \nonumber\\
&&m_d=-s\frac{S(C_4-S)}{CC_1+SC_4-1+\epsilon\delta CS}, \label{mud} \eea
while the Komar angular momenta $j_u$ and $j_d$ are defined by the
expressions
\bea
&&a_u\equiv\frac{j_u}{m_u}=s\frac{\epsilon\delta C[(C-\epsilon\delta S)C_1-1+\epsilon\delta CS]}
{(C_1+\epsilon\delta C_4)(CC_1+SC_4-1+\epsilon\delta CS)}, \nonumber\\
&&a_d\equiv\frac{j_d}{m_d}=s\frac{S[(S-\epsilon\delta C)C_4-1+\epsilon\delta CS]}
{(C_1+\epsilon\delta C_4)(CC_1+SC_4-1+\epsilon\delta CS)}, \label{aud} \eea
where the new constants $C$, $S$, $C_1$, $C_4$ and $\epsilon$ are
introduced via the relations
\be \varphi\equiv C+iS, \quad C_1\equiv\frac{1}{2}\epsilon_1\left(\omega_1+\frac{1}{\omega_1}\right),
\quad C_4\equiv\frac{1}{2}\epsilon_4\left(\omega_4+\frac{1}{\omega_4}\right),
\quad  \epsilon\equiv\epsilon_1\epsilon_4. \label{CS} \ee

The above Komar quantities (\ref{mud}) and (\ref{aud}) constitute
a set of four parameters with a clear physical meaning. Then a
question arises, whether these quantities can be used for
parametrizing the equilibrium solution? Remarkably, the answer is
yes, and the best practical way to do this is by means of the
following inversion formulas:
\bea &&C_1=C-\epsilon\delta\frac{m_u}{M+s}S, \nonumber\\
&&C_4=S-\epsilon\delta\frac{m_d}{M+s}C, \nonumber\\
&&C=\kappa\frac{M+s+\epsilon\delta a_u}{\sqrt{(M+s+\epsilon\delta a_u)^2+(M+s+\epsilon\delta a_d)^2}}, \nonumber\\
&&S=\kappa\frac{\epsilon\delta(M+s+\epsilon\delta a_d)}{\sqrt{(M+s+\epsilon\delta a_u)^2+(M+s+\epsilon\delta a_d)^2}}, \label{Tr2}
\eea
where $\kappa=\pm1$, while $s$ satisfies the quadratic equation
\be s^2+[2M+\epsilon\delta(a_u+a_d)]s+M^2+\epsilon\delta J=0, \quad M\equiv m_u+m_d, \quad J\equiv j_u+j_d. \label{bc} \ee
Note that Eq.~(\ref{bc}), after rewriting it in the form
\be \epsilon\delta(M+s)^2+s(a_u+a_d)+J=0, \label{be0} \ee
can be immediately recognized as the equilibrium law for two
arbitrary Kerr constituents originally derived in our paper
\cite{MRu2}.

Mention that the $\kappa$ sign, to be congruent with all our
previous conventions, has to be chosen in such a way that $C>0$.
It is also clear that, since $s>0$, the admissible values of the
masses and angular momenta are those that correspond to at least
one positive $s$ in Eq.~(\ref{bc}).

Therefore, the set $(z_0,m_u,m_d,j_u,j_d)$ can be used for
parametrizing the equilibrium class of the EDK solution.
Apparently, the constant $z_0$ can be always fixed at some
specific  value, for instance if one wants to bring the origin of
coordinates into the center of mass or into some other point
related to a concrete binary configuration that might look
attractive from the physical standpoint.

\section{Physical parametrization of $\a_i$ and $X_i$. The metric functions}

In order to rewrite the complex potential (\ref{E_pot}) and
corresponding metric (\ref{metric1}) in the physical parameters,
it is necessary to find the reparametrized form of the quantities
$\a_i$ and $X_i$. As it follows from (\ref{Tr1_inv}), the
constants $\a_i$ can be written in the form
\be \a_1=z_0+\frac{s}{2}+\s_u, \quad \a_2=z_0+\frac{s}{2}-\s_u,
\quad \a_3=z_0-\frac{s}{2}+\s_d, \quad \a_4=z_0-\frac{s}{2}-\s_d,
\label{alsi} \ee
where
\be
\s_u=s\frac{\delta\omega_4(\omega_1^2-1)}{(\omega_1+\delta\omega_4)(1+\delta\omega_1\omega_4)}, \quad
\s_d=s\frac{\omega_1(\omega_4^2-1)}{(\omega_1+\delta\omega_4)(1+\delta\omega_1\omega_4)}.
\label{sud1} \ee

The desired ``physical'' form of $\s_u$ and $\s_d$ is then
obtainable with the aid of formulas (\ref{CS})-(\ref{bc}),
yielding after tedious but straightforward algebraic manipulations
the following final expressions:
\bea
\s_u&=&\sqrt{m_u^2-a_u^2+m_da_u\frac{a_u(M+m_u+2s)-2m_u[a_d+\varepsilon(M+s)]}{(M+s)^2}}, \nonumber\\
\s_d&=&\sqrt{m_d^2-a_d^2+m_ua_d\frac{a_d(M+m_d+2s)-2m_d[a_u+\varepsilon(M+s)]}{(M+s)^2}},
\label{sud2} \eea
where $\varepsilon\equiv\epsilon\delta$. The above $\s_u$ and
$\s_d$ differ considerably from $\s=\sqrt{m^2-a^2}$ of a single
Kerr source \cite{Ker} due to interaction of the constituents. It
is worth mentioning that in the case of the real-valued $\a$'s,
say $\a_1$ and $\a_2$, the corresponding $\s_u^2>0$; however, if
$\a_2=\bar\a_1$ then $\s_u^2<0$ and one must use the convention
$\s_u=-i\sqrt{-\s_u^2}$ if one wants to pass to a positive
definite radicand in (\ref{sud2}). In Fig.~2 we have shown two
reparametrized equilibrium configurations for which the origin of
coordinates is chosen at the center of the lower constituent
($z_0=s/2$).

In a similar manner, by using (\ref{CS})-(\ref{bc}), it is
possible to rewrite formulas (\ref{Xi}) in terms of the Komar
quantities; below we give the resulting reparametrized form of
$X_i$:
\bea X_1&=&\frac{(M+s+\varepsilon a_d)(M+s-i\varepsilon
m_u)+i\varepsilon(M+s)\s_u} {(M+s+\varepsilon
a_d)(M+s+i\varepsilon m_u)-i\varepsilon(M+s)\s_u}, \nonumber\\
X_2&=&\frac{(M+s+\varepsilon a_d)(M+s-i\varepsilon
m_u)-i\varepsilon(M+s)\s_u} {(M+s+\varepsilon
a_d)(M+s+i\varepsilon m_u)+i\varepsilon(M+s)\s_u}, \nonumber\\
X_3&=&-\frac{(M+s+\varepsilon a_u)(M+s+i\varepsilon
m_d)+i\varepsilon(M+s)\s_d} {(M+s+\varepsilon
a_u)(M+s-i\varepsilon m_d)-i\varepsilon(M+s)\s_d}, \nonumber\\
X_4&=&-\frac{(M+s+\varepsilon a_u)(M+s+i\varepsilon
m_d)-i\varepsilon(M+s)\s_d} {(M+s+\varepsilon
a_u)(M+s-i\varepsilon m_d)+i\varepsilon(M+s)\s_d}.  \label{Xi1}
\eea

An alternative way of writing $X_i$ which may be advantageous for
some calculations is this:
\bea X_1&=&\frac{1}{m_u\Delta}[(M+s)(m_u+i\varepsilon\s_u)-\varepsilon(sa_u-m_ua_d)], \nonumber\\
X_2&=&\frac{1}{m_u\Delta}[(M+s)(m_u-i\varepsilon\s_u)-\varepsilon(sa_u-m_ua_d)],
\nonumber\\ X_3&=&\frac{i}{m_d\Delta}[(M+s)(\varepsilon
m_d+i\s_d)-sa_d+m_da_u)], \nonumber\\
X_4&=&\frac{i}{m_d\Delta}[(M+s)(\varepsilon
m_d-i\s_d)-sa_d+m_da_u)], \nonumber\\ \Delta&\equiv&
-(M+s+\varepsilon a_u)+i[\varepsilon(M+s)+a_d]. \label{Xi2} \eea

Now we are able to write down the reparametrized complex potential
(\ref{E_pot})-(\ref{Xi}); its new simple representation is the
following:
\bea \E&=&E_-/E_+, \nonumber\\
E_\mp&=&[s^2-(\s_u+\s_d)^2](X_1r_1-X_2r_2\mp2\s_u)(X_3r_3-X_4r_4\mp2\s_d)
\nonumber\\
&&-4\s_u\s_d[X_2r_2-X_3r_3\mp(s-\s_u-\s_d)][X_1r_1-X_4r_4\mp(s+\s_u+\s_d)],
\nonumber\\ r_i&=&\sqrt{\rho^2+(z-\a_i)^2}, \nonumber\\
\a_1&=&\frac{s}{2}+\s_u, \quad \a_2=\frac{s}{2}-\s_u, \quad
\a_3=-\frac{s}{2}+\s_d, \quad \a_4=-\frac{s}{2}-\s_d,
\label{E_sim} \eea
where $\s_u$, $\s_d$ and $X_i$ are determined by (\ref{sud2}) and
(\ref{Xi1}) or (\ref{Xi2}), and where we have set $z_0=0$ in the
expressions of $\a$'s.

The above potential $\E$ can be also written in the form
\bea \E&=&\frac{\Lambda+\Gamma}{\Lambda-\Gamma}, \nonumber\\
\Lambda&=&[s^2-(\s_u+\s_d)^2](X_1r_1-X_2r_2)(X_3r_3-X_4r_4)
-4\s_u\s_d(X_2r_2-X_3r_3)(X_1r_1-X_4r_4), \nonumber\\
\Gamma&=&2\s_d\{[(s+\s_u)^2-\s_d^2]X_2r_2-[(s-\s_u)^2-\s_d^2]X_1r_1\}
\nonumber\\
&&+2\s_u\{[(s-\s_d)^2-\s_u^2]X_4r_4-[(s+\s_d)^2-\s_u^2]X_3r_3\},
\label{E_pot2} \eea
and below we will use the functions $\Lambda$ and $\Gamma$ for
presenting the reparametrized coefficients $f$, $\gamma$ and
$\omega$ in the metric (\ref{metric1}):
\bea
f&=&\frac{\Lambda\bar\Lambda-\Gamma\bar\Gamma}{(\Lambda-\Gamma)(\bar\Lambda-\bar\Gamma)},
\quad  e^{2\gamma}=\frac{\Lambda\bar\Lambda-\Gamma\bar\Gamma}{K_0
r_1r_2r_3r_4}, \quad \omega=\omega_0 -\frac{2{\rm
Im}[(G(\bar\Lambda-\bar\Gamma)]}{\Lambda\bar\Lambda-\Gamma\bar\Gamma},
\nonumber\\
G&=&z\Gamma+4s\s_u\s_d[(X_3r_3+\a_3)(X_4r_4+\a_4)-(X_1r_1+\a_1)(X_2r_2+\a_2)]
\nonumber\\
&&+(\s_u+\s_d)[s^2-(\s_u-\s_d)^2][(X_1r_1+\a_1)(X_3r_3+\a_3)-(X_2r_2+\a_2)(X_4r_4+\a_4)]
\nonumber\\
&&+(\s_u-\s_d)[s^2-(\s_u+\s_d)^2][(X_2r_2+\a_2)(X_3r_3+\a_3)-(X_1r_1+\a_1)(X_4r_4+\a_4)],
\nonumber\\ K_0&=&\frac{64}{m_1^2m_2^2}s^2|\s_u|^2|\s_d|^2(M+s)^2,
\quad \omega_0=-2\varepsilon(M+s). \label{metric2} \eea

Therefore, we have obtained a physical representation for the
general family of equilibrium configurations in the EDK solution.
Its interesting particular case which we would like to mention in
conclusion of this section is the DH configuration for two
balancing identical corotating superextreme Kerr particles
\cite{DHo} possessing an additional symmetry with respect to the
equatorial plane \cite{Kor,MNe}. For this specific two-body system
$m_u=m_d=m$, $a_u=a_d=a$, $\s_u=\s_d=\s$, and it is convenient to
solve Eq.~(\ref{bc}) for $a$, yielding ($\delta=+1$)
\be a=-\frac{\epsilon(s+2m)^2}{2(s+m)}, \label{a_es} \ee
which means that $m$ and $s$ are chosen as arbitrary parameters of
the solution. Then we readily obtain for $X_i$ the expressions
\bea
X_1&=&\frac{s+(2-i\epsilon)m+\epsilon\mu}{s+(2+i\epsilon)m-\epsilon\mu},
\quad
X_2=\frac{i\epsilon[s+(2+i\epsilon)m-\epsilon\mu]}{s+(2-i\epsilon)m+\epsilon\mu},
\nonumber\\
X_3&=&\frac{i\epsilon[s+(2-i\epsilon)m+\epsilon\mu]}{s+(2+i\epsilon)m-\epsilon\mu},
\quad
X_4=-\frac{s+(2+i\epsilon)m-\epsilon\mu}{s+(2-i\epsilon)m+\epsilon\mu},
\label{Xi_es} \eea
while $\s$ becomes a pure imaginary quantity (since $m>0$, $s>0$)
whose explicit form is the following:
\be \s=-\frac{is\mu}{2(s+m)}, \quad \mu\equiv\sqrt{s^2+6ms+7m^2}.
\label{s_es} \ee
For $\a_i$ and $r_i$ in the equatorially symmetric case one has
\bea &&\a_1=-\a_4=\frac{s}{2}+\s, \quad
\a_2=-\a_3=\frac{s}{2}-\s, \nonumber\\
&&r_1=\sqrt{\rho^2+(z-\a_1)^2}, \quad
r_2=\sqrt{\rho^2+(z-\a_2)^2}, \nonumber\\
&&r_3=\sqrt{\rho^2+(z+\a_2)^2}, \quad
r_4=\sqrt{\rho^2+(z+\a_1)^2}, \label{al_se} \eea
and the potential $\E$ of the DH equilibrium configuration, after
the substitutions into formulas (\ref{E_pot2}) and subsequent
simplifications, finally takes the form
\bea && \E=\frac{\Lambda+\Gamma}{\Lambda-\Gamma}, \nonumber\\ &&
\Lambda=(s^2-4\s^2)(\mu_-r_2-\mu_+r_1)(\mu_+r_3-\mu_-r_4)
-4\s^2(\mu_-r_2-i\epsilon\mu_+r_3)(i\epsilon\mu_+r_1+\mu_-r_4),
\nonumber\\ &&
\Gamma=2ms\s[(1-i\epsilon)(s-2\s)(\mu_-r_4+i\epsilon\mu_+r_1)
-(1+i\epsilon)(s+2\s)(\mu_-r_2-i\epsilon\mu_+r_3)], \label{E_DH}
\eea
whereas the corresponding metric functions $f$, $\gamma$ and
$\omega$ can be written as
\bea
f&=&\frac{\Lambda\bar\Lambda-\Gamma\bar\Gamma}{(\Lambda-\Gamma)(\bar\Lambda-\bar\Gamma)},
\quad  e^{2\gamma}=\frac{\Lambda\bar\Lambda-\Gamma\bar\Gamma}{K_0
r_1r_2r_3r_4}, \quad \omega=\omega_0 -\frac{2{\rm
Im}[(G(\bar\Lambda-\bar\Gamma)]}{\Lambda\bar\Lambda-\Gamma\bar\Gamma},
\nonumber\\
G&=&z\Gamma+s\s\{2s(\mu_-^2r_2r_4-\mu_+^2r_1r_3)-8i\epsilon
m^2\s(r_1r_2+r_3r_4)+(1-i\epsilon)m(s^2-4\s^2) \nonumber\\
&&\times[\mu_+(r_3+i\epsilon r_1)-\mu_-(r_4+i\epsilon r_2)]\},
\nonumber\\  K_0&=&256s^2\s^4(s+2m)^2, \quad
\omega_0=-2\epsilon(s+2m), \quad \mu_\pm\equiv s+3m\pm\epsilon\mu.
\label{met_DH} \eea

To consider a particular DH configuration, one only needs to
choose the values of $m$ and $s$, and find from (\ref{a_es}) the
corresponding value of $a$ at which the balance occurs. Formulas
(\ref{s_es})-(\ref{met_DH}) will then describe the spacetime for
that parameter choice.

\section{Discussion}

Although the general formulas worked out in the previous section
are applicable to all four types of the two-Kerr configurations
from Fig.~1, the equilibrium states with $m_u>0$, $m_d>0$ are only
possible for the systems ($b$), ($c$) and ($d$) containing at
least one superextreme component. Various particular equilibrium
configurations between a black-hole and a superextreme
constituents, or between two unequal superextreme constituents
were considered in the paper \cite{MRS}, and recently we have
shown \cite{MRu3} that balance can be achieved even between a
Schwarzschild black hole and a Kerr superextreme object. The
absence of the equilibrium between two underextreme Kerr
constituents with positive Komar masses (the systems (a) in
Fig.~1) was strictly proved in our paper \cite{MRu1}, and the
non-existence proof was later extended to the case of two extreme
Kerr constituents \cite{CMR}, thus ruling out the two-black-hole
equilibrium states in the EDK solution.

Remarkably, the expressions for the areas of the horizons
calculated for the equilibrium configurations of type ($a$) with
the aid of Tomimatsu's formulas \cite{Tom2}
\be A_u=2\pi(\a_1-\a_2)\sqrt{-\omega_ue^{2\gamma_u}}, \quad
A_d=2\pi(\a_3-\a_4)\sqrt{-\omega_de^{2\gamma_d}}, \label{ah} \ee
where $\omega_u$, $\omega_d$, $\gamma_u$, $\gamma_d$ are constant
values of the functions $\omega$ and $\gamma$ on the respective
horizons, are able to provide us with a simple demonstration that
the individual Komar masses $m_u$ and $m_d$ cannot simultaneously
take on positive values in such configurations. Taking into
account that $\delta=+1$ in the ($a$)-type equilibrium states, one
can arrive at the following final expressions for $A_u$ and $A_d$:
\bea A_u=-\frac{4\pi m_u[(s+m_d)(M+s+\epsilon
a_d)-\s_u(M+s)]^2}{s(M+s)(M+s+\epsilon a_d)}, \nonumber\\
A_d=-\frac{4\pi m_d[(s+m_u)(M+s+\epsilon
a_u)-\s_d(M+s)]^2}{s(M+s)(M+s+\epsilon a_u)}, \label{ahf} \eea
whence it follows immediately that in order the masses of the
black-hole constituents and areas of the horizons could take
positive values simultaneously, the following two conditions must
be satisfied:
\be M+s+\epsilon a_d<0, \quad M+s+\epsilon a_u<0. \label{condA} \ee
However, after rewriting the equilibrium condition (\ref{bc}) in
the form ($\delta=+1$)
\be s(M+s)-(m_u+s)(M+s+\epsilon a_u)-(m_d+s)(M+s+\epsilon a_d)=0,
\label{bc_re} \ee
we see that, under the suppositions made, the inequalities
(\ref{condA}) convert the left-hand side of (\ref{bc_re}) into a
strictly positive quantity, which signifies the absence of
equilibrium configurations of two Kerr black holes. Note, however,
that in the systems ($b$) and ($c$) the black-hole component has
the horizon area defined by one of the expressions (\ref{ahf}),
with $\epsilon$ substituted by $\epsilon\delta$, so that the
balance condition (\ref{bc_re}) may have physically meaningful
solutions because in such systems only one of the inequalities
(\ref{condA}) has to be satisfied.

It would certainly be of interest to briefly discuss a direct
mathematical interrelation existing between the equilibrium
configurations of the EDK solution and the analogous
configurations of the double-Reissner-Nordstr\"om (DRN) solution
\cite{Man,BMA}. While the former configurations are defined by the
condition (\ref{be0}), the latter equilibrium states of two
electrically charged Reissner-Nordstr\"om sources \cite{Rei,Nor}
are defined by the balance condition
\be m_um_d-\left(q_u+\frac{m_uq_d-m_dq_u}{m_u+m_d+s}\right)
\left(q_d+\frac{m_dq_u-m_uq_d}{m_u+m_d+s}\right)=0 \label{beDRN}
\ee
(the reader is referred to \cite{Man,BMA,ABe} for the details of
its derivation), where $m_u$ and $m_d$ are Komar masses of the
upper and lower constituents, $q_u$ and $q_d$ are the
corresponding charges, while s is the relative coordinate
distance. The connection between Eqs.~(\ref{be0}) and
(\ref{beDRN}) is described by the following two theorems.

\medskip

\noindent{\it Theorem I.} If $m_u$, $m_d$, $a_u$, $a_d$, $s$ is an
equilibrium configuration of the EDK solution, then the
substitution
\be a_u=\frac{\epsilon\delta q_u(m_uq_d-m_dq_u)}{m_um_d-q_uq_d},
\quad a_d=\frac{\epsilon\delta q_d(m_dq_u-m_uq_d)}{m_um_d-q_uq_d},
\quad m_dq_u-m_uq_d\ne0, \label{sub1} \ee
into Eq.~(\ref{be0}) defines an equilibrium configuration of the
DRN solution.

\medskip

\noindent{\it Theorem II.} Given an equilibrium configuration of
the DRN solution, $m_u$, $m_d$, $q_u$, $q_d$, $s$, the
substitution
\be q_u^2=-\frac{m_um_da_u^2}{\Delta_0}, \quad
q_d^2=-\frac{m_um_da_d^2}{\Delta_0}, \quad
q_uq_d=\frac{m_um_da_ua_d}{\Delta_0}, \label{sub2} \ee
with $\Delta_0\equiv a_ua_d+\epsilon\delta(m_ua_d+m_da_u)\ne0$,
$\Delta_0 m_um_d<0$, converts Eq.~(\ref{beDRN}) into condition
(\ref{be0}).

\medskip

The proof of these theorems is straightforward and consists in the
substitution of (\ref{sub1}) and (\ref{sub2}) into
Eqs.~(\ref{be0}) and (\ref{beDRN}), respectively.

\section{Conclusion}

We hope that the physical representation of the general family of
equilibrium configurations of two Kerr sources obtained in the
present paper will make this family more accessible for concrete
applications and will simplify the analysis of particular cases
which exhibit interesting physical properties. Although two Kerr
black holes cannot be in the gravitational equilibrium, this fact
does not diminish the importance of the EDK solution because there
are other physically meaningful equilibrium configurations it
offers -- those between a black hole and a superextreme source,
and between two superextreme Kerr constituents, both types of the
configurations permitting their components to have exclusively
positive Komar masses. It is probably worth remarking that for
many years the superextreme solutions had been largely
underestimated compared to the black-hole ones in spite of the
theoretical evidence that they may arise from the gravitational
collapse \cite{GJVW,Jos}, or are able to open new horizons for the
gravitational experiment (an important prediction made four
decades ago by Penrose \cite{Pen}). In relation with the latter
aspect we would like to emphasize that the discovery of the
physically relevant equilibrium states between a black-hole and a
superextreme Kerr constituents (for particular examples we refer
the reader to \cite{MRS}) is highly important from the physical
point of view, mainly because the balance in such two-body systems
might signify that the spin-spin repulsive force is actually by
far stronger than was thought in the 1980's when only the
equilibrium configurations composed of two superextreme objects
were found, and in our opinion this could have relevance to the
experimental detection of the spin-spin interaction. It also
appears that the recent paper of Jacobson and Sotiriou \cite{JSo}
on destroying black holes with test bodies establishes an
interesting physical bridge between the two types of exact
solutions, and we expect that the binary equilibrium
configurations described by the EDK solution will be able to shed
additional light on the physical interaction of black holes and
superextreme sources.

\section*{Acknowledgements}
We thank the referees for useful comments and suggestions. This
work was partially supported by CONACYT, Mexico, and by Ministerio
de Ciencia y Tecnolog\'\i a, Spain, under the Projects
FIS2009-07238 and FIS2012-30926.

\newpage

\begin{figure}[htb]
\centerline{\epsfysize=90mm\epsffile{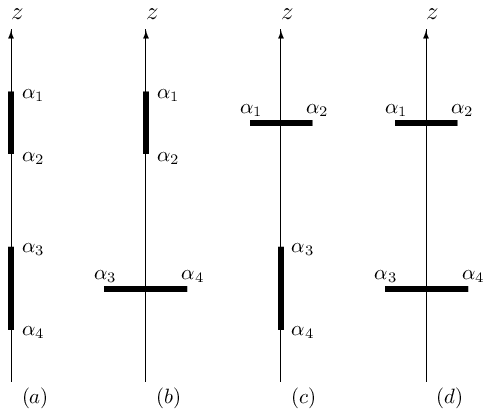}} \caption{Four
different types of the equilibrium configurations of two Kerr
sources.}
\end{figure}

\begin{figure}[htb]
\centerline{\epsfysize=90mm\epsffile{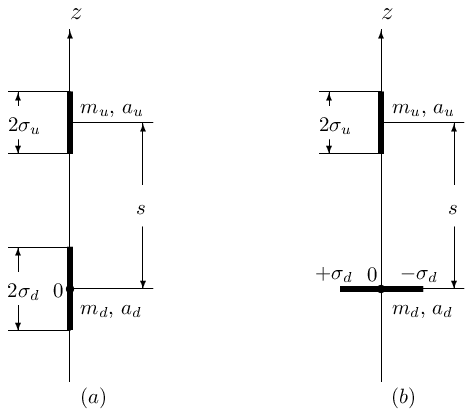}} \caption{Physical
reparametrization of the equilibrium configurations ($a$) and
($b$) from Fig.~1.}
\end{figure}

\end{document}